# The impact of financial risks on economic growth in EU-15

**Ionuț JIANU**
The Bucharest University of Economic Studies, Romania
ionutjianu91@yahoo.com
**Laura-Mădălina PÎRȘCOVEANU**
The Bucharest University of Economic Studies, Romania
laura.pirscoveanu@yahoo.com
**Maria-Daniela TUDORACHE**
The Bucharest University of Economic Studies, Romania
mariadanielatudorache@gmail.com

**Abstract.** *This paper examines the impact of financial risks on economic growth in the first 15 Member States of the European Union, considering 1995-2014 period and aims to lay down a new explanatory model of economic growth, based mainly on the behavioral reactivity of the financial disruptions mentioned above. The model was estimated through the panel estimated generalized least squares method and included additional control variables in order to strengthen the research conducted. Our goal consists in the examination of the financial risks in the European Union and in the estimation of their impact on economic growth.*

**1. Introduction**

The financial crisis was the biggest challenge for the global economy and European Union (EU), from the postwar period to the present. The economic slowdown was caused mainly by the disturbances manifested on the credit markets from the European Union and by the losses incurred by the systemically important credit institutions.

Thereby, the confidence in the financial system to meet payment obligations crashed and financial, economic and political risks stood on an upward trend and experienced a high volatility during this period. These kind of risks has distorted economic growth and drew the premises of building an economy of fear.

Financial risks arise from the likelihood of incurring financial costs as a result of the investments placed in companies which are in default on their financial obligations.

Subsequently, national governments have guaranteed for the banking sector debt through government bonds, in order to save the sector, because some credit institutions were too important to be left to fail. Thus, bank debt was converted into public debt and high levels of it has led to additional financial risks.

Our motivation in choosing this subject lies in the great importance of financial risks evolution, regarding the influence exerted by them on future economic cycles. The growth of financial risks reduce the investors confidence to place funds in certain economic activities and also, may impact the country rating in a negative sense.

The main goal of the paper is to analyse the impact of financial risks on economic growth in the first 15 member states of the European Union, taking into account the following specific objectives:
- performing the comparative analysis of financial risks at aggregate and granular level in the EU-15 (1995-2014 period);
- estimating the impact of financial risks on economic growth in the EU-15;
- identifying other relevant determinant factors for the endogenous variable and estimating their coefficients;
- testing the hypotheses of the model in order to confirm its validity.

In this paper, we chose to analyse the countries belonging to the EU-15 group, because in 1995 (starting point of the analysis), the EU included 15 Member States.

**2. Literature review**

The economic development and growth are the main goals that all countries should take into consideration in developing national strategies. The financial crisis started in 2007 in the United States and has affected EU countries in less than a year, 2008 and 2009 representing a difficult period when economies and finances of Member States were affected. After a slight improvement in 2010, the pessimistic scenarios of economists regarding the persistence of the recession have become true. Blyth (2013) noted that recent developments in the sovereign debt crisis is a result of the financial crisis, as governments have intervened to save the banking sectors and, therefore the financial crisis has turned into a debt crisis. All of this led to an increase in financial risks that have affected economic growth. The growth of disturbances mentioned-above grew the financial system vulnerability and the whole system became prone to shocks.

Dinu (2011) asserted that the signalised trends do not coincide with those existing on the market, this condition being caused by the estimation of neglected, erroneous and falsified signals by the economists community as a result of too objective and mechanical thinking in testing economic hypotheses.

The PRS Group computes the financial risk, placing the indicator values on a scalar field, following the aggregation of its subcomponents, which are represented by the following risks: current account, external debt, external debt service, exchange rate stability and international liquidity. Next, we specified a number of authors who have analysed the impact of financial risk subcomponents on economic growth, respectively their evolution in the European Union.

Regarding the evolution of the current account, Jaumotte and Sodsriwiboon (2010) analysed the current account imbalances in the southern Eurozone sub-model and identified monetary integration (process which has greatly increased access to foreign savings) as the main cause of current account balances reduction. Holinski et al. (2012) demonstrated that a large share of the current account surplus registered in the northern sub-model of euro zone is due to fiscal consolidation, savings, relatively stable private investments, competitiveness and higher revenues recorded abroad. At the opposite, southern sub-model has experienced a reduction in private savings since 1992, which led to loan growth and trade balance deficits. The authors believe that excessive risk-taking by the banking sector and the procyclicality of the common monetary policy in the euro area are possible determinants of the current account imbalances, ex-ante crisis. Socol (2011) demonstrated that the current account deficit occurs due to lower domestic production and budget deficits, considering the fact that, if a country goes into recession, it will require additional budget expenditures. However, in that country, the price level will decrease due to the aggregate demand reduction. Nevertheless, the decrease level depends on the elasticity of the aggregate supply curve. On the other hand, a current account surplus implies an increase in production and exports, resulting in a budget surplus caused by the production growth. Smith and Lazarus (2015) identified a positive impact of higher current account balance on economic growth in advanced countries in the European Union.

Pattillo et al. (2011) analysed the relationship between external debt and economic growth in 93 emerging countries and identified a negative impact of debt on gross domestic product (GDP) growth in countries with an average level of the endogenous. According to their research, the debt doubling from the countries with an average foreign debt reduce economic growth by one third or a quarter of the previously increase. The authors identified a negative impact of external debt on growth, in the case of debt level exceeding the 160-170% threshold of total exports of goods and services or 35-40% of GDP. The rising of external debt diminish the investors' expectations regarding their financial investments yields, anticipating a tax rates increment. This instability, finally, has the ability to blur technological progress and the efficient allocation of resources. Also, a high level of external debt diminishes the ability of states with low income to provide social services such as health and education. Randveer et al. (2011) analysed the effects of private debt on economic growth in 31 OECD countries and 20 emerging countries. The authors found a high correlation between the high levels of debt (before entering into a recessionary phase) and low growth rates, ex-post recession. Regarding public debt, Reinhart et al. (2010) found that a high proportion of public debt to GDP leads to a decline in economic growth. On the other hand, Dreger and Reimers (2013) showed that the negative impact of public debt increase (expressed as a percentage of GDP) on growth is specific to the euro area and to the periods when there are recorded unsustainable debt levels. Egert (2015) identified a negative nonlinear relationship between the government debt and economic growth, which is influenced by the analysed sample, data frequency, as well as, by the method applied.

Clements et al. (2003) analysed the impact of external debt service on growth in low-income countries and discovered that the exogenous variable decrease affects indirectly (through its impact on public investments) and negatively the economic growth. The negative effect of higher external debt service on growth was also confirmed by Kohlscheen (2008). Gohar et al. (2009) investigated the impact of external debt service on growth in low-income countries, using the panel method and obtained a negative and significant coefficient of the estimator. Also, Shabbir (2013) examined the influence of the increase of the external debt service on the growth rate of gross national income and concluded that the impact is negative. Mihuț and Câlea (2015) proved that in the EU there is a negative relationship between external debt and economic growth. The growth of the interest expenses increase the budget deficit, this having the ability to have adverse consequences on growth. At the same time, a high level of payments of principal repayments, interest and commissions are a source of concern for investors and an increase in these indicators may reduce the economic sentiment.

Regarding the relationship between exchange rate stability and growth, finding a direct connection is a challenge. Grauwe and Schnabl (2004) demonstrated that the fixed exchange rate regime stimulates international trade and division of labour, which leads to interest rate decrease, thereby stimulating investment and economic growth. On the other hand, Huang and Malhotra (2005) found that the choice of exchange rate regime does not have a significant impact on economic growth in European countries, but the adoption of a flexible exchange rate regime has the ability to influence the level of economic growth. Other authors have argued the importance of exchange rate stability to maintain a steady pace of growth. Broda and Romalis (2011) highlighted that exchange rate volatility reduce the differentiated goods trade. Also, Rapetti (2013) demonstrated that a high fluctuation of the real exchange rate (measured generally by the standard deviation of the variation coefficient of the real exchange rate) affects economic growth, adversely.

Polterovich and Popov (2002), Cruz and Kriesler (2008) and Levy-Yeyati and Sturzenegger (2009) identified a positive impact of international reserves on growth.

International liquidity becomes an element of utmost importance during periods when economic growth adopted measures are not feasible by political or any other reasons. For example, a country politically unstable, with a precarious social welfare, can have an effective economic development through the accumulation of foreign reserves by the central bank. Bussière et al. (2014) analysed the usefulness of official international reserves, stating that the countries who have held a high share of reserves in the short-term debt had a great resilience to the financial crisis shock. Krushkova and Maric (2015) confirmed the positive effect on growth, as a consequence of international reserves accumulation, using a panel data model for countries such as Brazil, China and Russia during 1993-2012 period. Also, Cheng (2015) found a high correlation between international official reserve increase and economic growth in emerging countries.

### 3. Methodology

This section describes the methodology and techniques applied in order to perform the empirical analysis. Combining qualitative and quantitative approach, using the deductive method in both cases was essential to provide an additional level of robustness to our assessment.

Investigating the impact of financial risks on economic growth in the EU-15 requires, in a first phase, analysing the regressors evolution at granular level. In this regard, we extracted the indicators related to financial risks, as well as its sub-indicators from PRS Group platform for the first 15 EU Member States, the series of data covering the 1995-2014 period. Given that the frequency of the data is monthly and our study is based on annual data, we computed the annual average for each indicator. The PRS Group scales the financial risk indicators within the following ranges:
- financial risk (0-50, 50 being the lowest risk);
- risk for current account as a share of total exports of goods and services (0-15, 15 being the lowest risk);
- risk for debt service (0-10, 10 being the lowest risk);
- risk for exchange rate stability (0-10, 10 being the lowest risk);
- risk for external debt (0-10, 10 being the lowest risk);
- risk for international liquidity (0-5, 5 being the lowest risk).

Our research involved performing the risk analysis on two sub-periods through the use of comparative analysis as a research method, the sub-periods being separated, according to the moment when we found a change of trend, one of them including the financial crisis. We also calculated the standard deviation of the risks, this having the role to be a proxy for the financial risks volatility (a high level of it highlighting the instability of the economy) manifested during the 1995-2014 period, using the formula:

$$SD = \sqrt{\frac{\sum_{i=1}^{n}(x_i-\bar{x})^2}{n-1}} \qquad (1)$$

Regarding quantitative approach, using Eviews 9.0, we estimated the impact of financial risks (accompanied by other control variables) on growth in the EU-15 by panel estimated generalized least squares fixed effect model (weighted by Cross-section SUR option), data being extracted with annual frequency for the 1995-2014 period from platforms such as: AMECO, Eurostat, PRS Group and World Bank (Annex 1). We used Cross-section SUR option in order to correct the possible inconveniences of the model, related to the presence of the heteroscedasticity and that of the autocorrelation between cross-sections. The observations (300 for 15 cross-sections and 285 observations after adjustments) included in the analysis increased the ability of the used method to give a greater efficiency of the following model:

$$y_t = \alpha_0 + \beta_0 y_{t-1} + \beta_1 imp_t + \beta_2 npt_t + \beta_3 budget_{t-1} + \beta_4 intrate_{t-1} +$$
$$+ \beta_4 inflation_{t-1} + \beta_5 finrisk_{t-1} + \beta_6 yrisk_{t-1} + \varepsilon_t \qquad (2)$$

where:

$y_t, y_{t-1}, imp_t$, respectively $npt_t$ surprise the evolution of the economic growth and its autoregressive term, as well as that of the imports and net taxes applied to products, while $budget_{t-1}, intrate_{t-1}, inflation_{t-1}, finrisk_{t-1}$ and $yrisk_{t-1}$ represents the budgetary balance, the real long-term interest rate, inflation rate and the financial risks related to economic growth, all of these being lagged by 1 year, according to the economic theory.

In order to confirm the methodology used, we tested the stationarity for each variable through the test „Summary" which aggregates the results of the following tests: Levin, Lin and Chu, Breitung, Im, Pesaran and Shin, the ADF Fisher, respectively PP-Fisher. In exceptional cases, we used the statistical correlogram for identifying the presence or absence of trend, a situation that helped us establishing the initial condition of the following hypotheses: the presence of trend and constant, the presence of constant and the absence of trend and constant.

In some cases, the panel technique eliminates the problems related to the non-stationary character of the variables, but it brings new challenges for the heterogeneity of the data set. For this purpose, we used both Hausman and Redundant fixed effects – likelihood ratio tests, in order to choose the optimal effects estimation method between the random effects and the fixed effects methods. After the selection of the most appropriate method for estimating the coefficients, we checked the model validation hypotheses using the following tests:

- verifying the validity of the model (F-statistic);
- testing the residuals distribution (Jarque-Berra);
- examination of the residuals autocorrelation and cross-section dependence (Durbin Watson, Breusch-Pagan and Pesaran CD);
- testing homoscedasticity (White test) – we compared the product of the coefficient of determination and the number of observations with Chi-square statistic; if Chi-square has a higher value than the product, it can be confirmed the homoscedasticity hypothesis of the model;
- checking for multicollinearity (Klein's criterion) – Klein's criterion involves comparing the statistical correlation Pearson between the exogenous variables with the coefficient of determination. If the statistical correlation between two exogenous variables is greater than the coefficient of determination, then the presence of the multicollinearity can be accepted. Otherwise, multicollinearity absence is confirmed.

## 4. Results and interpretations

Depending on the methodology assumed, in the first phase, we analysed the evolution of the financial risks in EU-15 at aggregate and granular level.

According to the statistics published by the PRS Group, the lowest financial risks were manifested in 1996 (Figure 1), while in 2012 it was recorded the peak of the financial risks, a situation influenced mainly by the unfavorable evolution of the risks for external debt and international liquidity. The minimum point of the risk was based, primarily on the favorable evolution of the current account as a percentage of total exports of goods and services. Also, in 2005, after a slight improvement of the financial risk until 2004, there was a change of trend, which made us to separate the analysis in two equally sized sub-periods (1995-2004 and 2005-2014). This change of trend was caused by the evolution of the risk for foreign debt, a driver of the financial risk that felt an excessive increase from 2005 to the present.

First, our analysis consisted in the individual investigation of the EU-15 financial risks, as well as the analysis of its subcomponents, including the interpretation of the financial components evolution (Figure 2) and that of the risks recorded in the first and second period (Figure 3).

**Figure 1.** *The evolution of financial risks in the EU-15*

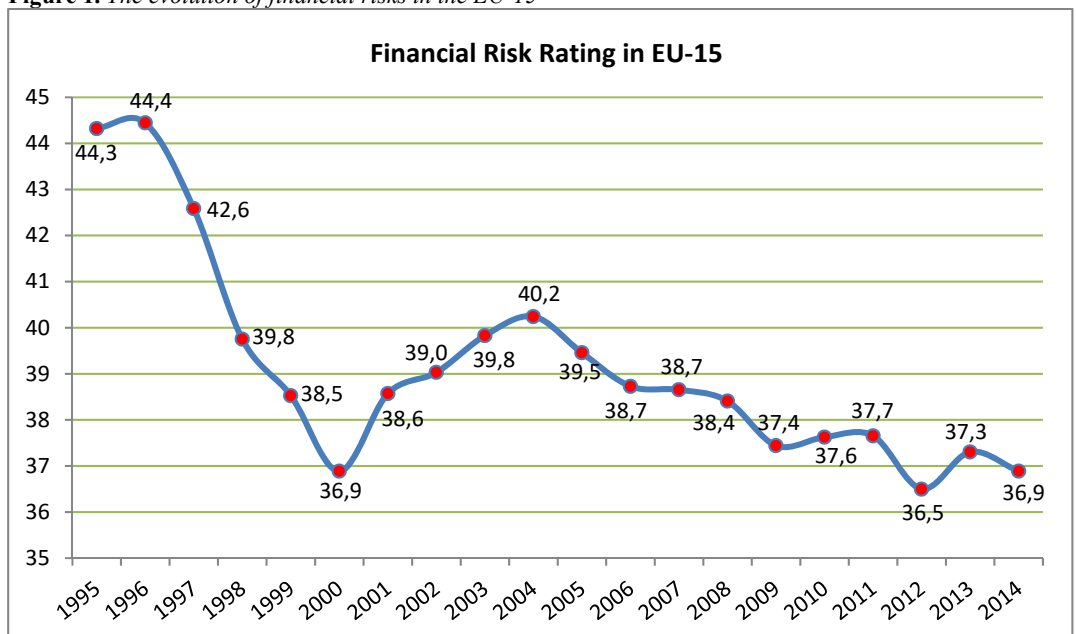

**Source:** Own calculations using PRS Group database.

**The financial risk.** We found that the highest growth of financial risks (compared to the value recorded in the previous year) occurred in 1998 (6.65%), while in 2013 it was recorded the largest decrease of the financial risks, post-crisis (2.21%), being preceded in 2012, as a consequence of high risks for external debt, by the highest percentage change of the 2005-2014 period (+ 3.07%).

**Figure 2.** *The evolution of financial risks components in the EU-15*

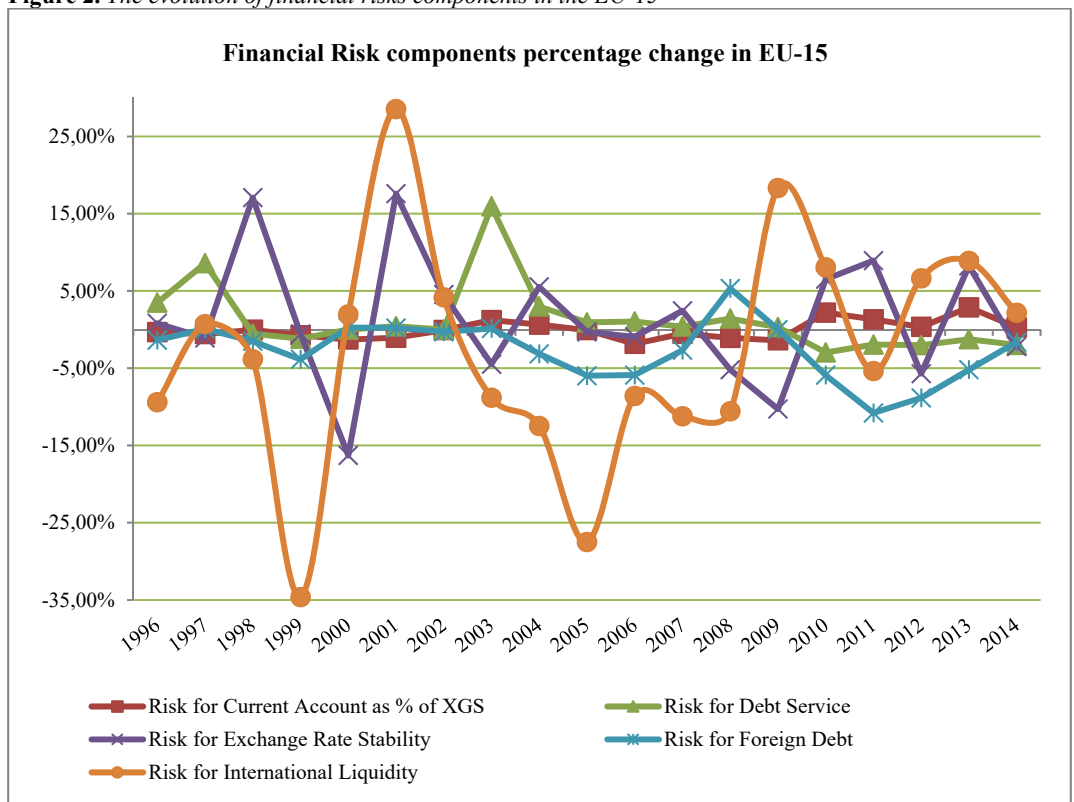

**Source:** Own calculations using PRS Group database.

Moreover, the decrease of financial risks in 2001 with 4.56% (the largest fall of 1995-2014) annulled the financial risk increase (4.25%) manifested in 2000, a situation caused mainly by the high fluctuation of the exchange rate in 2000. The reason why the USD/EUR exchange rate volatility reached high levels in 2000 compared to 2001 consists in the world perception towards the euro. In 1999, the euro was introduced as a virtual currency, following to be inserted on market since 2002. In 2000, the euro depreciated with 12.97% in December against the USD, while the depreciation of the euro in 2001 occurred in a lower rhythm, the USD/EUR exchange rate rising from 1.066 in January to 1.121 in December (5,16% depreciation of the euro). The explanation of this divergent evolution from the point of view of exchange rate stability consists in the low level of the confidence in the currency, manifested in 2000 (before euro introduction as a currency on market), which led to a strong depreciation of the euro and to a perception improvement towards the currency, due to the appropriation of the euro introduction on market.

Although, some studies have analysed the impact of the referendum organized in Denmark for abolishing the opt-out clause (28 September 2000), the impact of this event on the depreciation of the euro was only temporary (2 months), the USD/EUR exchange rate rising from 1.147 in September 2000 to 1.168 in November 2000, following to decrease at the end of December to 1.115.

EU-15 felt an increase of financial risks with 6.31% in the period 2005-2014, compared to 1995-2004. In the first examined period, Greece, Portugal and Sweden had the highest level of financial risks, while Luxembourg, Austria and Denmark were at the opposite pole. In the second analysed period, Sweden and Denmark were placed better from the perspective of the indicator evolution, these countries representing the only states from EU-15 that have managed a decrease of financial risks during this period, compared to the 1995-2004 period (a decrease by 1.55 deviation points for Sweden, respectively 0.79 deviation points for Denmark), Spain taking the position of Sweden in the category of countries with excessive financial risks, alongside Greece and Portugal. On the other hand, the most performing countries in the second period were Denmark, Germany and Luxembourg. We notice that, in countries like Ireland, Austria and Spain, in the second period, it have been manifested the most significant increases of financial risks, compared to the previous period, the indicator level decreasing with 6.36, 5.57 and 4.80 deviation points.

Denmark and Sweden have managed to diverge from the financial risks growth trend of EU-15 through a better position of the current account in the second period, as well as, due to significant reductions of the risks related to external debt service, exchange rate stability and international liquidity. At the opposite pole, was situated Ireland, Austria and Spain, the increasing of financial risks being caused, largely, by the growth of external debt risk and the international liquidity risk in all three cases, and by the increase of the risks for current account and external debt service in the case of Ireland, respectively by the increase of the risk for current account in the case of Spain.

On the other hand, Luxembourg and Germany are among the most advanced 3 member states from EU-15, from the point of view of the level of financial risks recorded during the 2005-2014 period. Germany reduced its financial risks related to current account, debt service and exchange rate stability, while Luxembourg has oriented towards the decrease of the risks for external debt service and exchange rate stability. Greece and Portugal remained among the last countries in the EU-15 in the second analysed period, recording an increase of risks for current account, external debt and international liquidity. In this context, Greece and Portugal had a similar trend of financial risks modification in 2005-2014, compared to 1995-2004, with the one manifested in Germany, excepting the evolution of the risk for current account, this one having an unfavorable position in the two peripheral countries, respectively favorable in Germany. Therefore, we found that one of the main reasons that made the difference regarding financial risks recorded in EU-15 consisted in the evolution of the current account. Specifically, Portugal and Greece have become market outlets for Germany and favored the growth prospects of its economy. According to Eurostat, in the 2005-2014 period (compared to the previous one), the exports from Germany to Greece increased by 37.91%, while the exports from Germany to Portugal grew by 27.09%, which supports our assumption.

**The risk for current account (as a percentage of total exports of goods and services).** For the time period analysed, the minimum level of risk for current account was registered in 1995, while in 2009, starting with financial crisis, we found the highest risk for current account, year that recorded the highest percentage increase of the risk post-crisis (1.39%).

In 2006 took place the highest increase of the risk for current account (1.84%), that being influenced by the exports growth below potential of Euro Area-12 (EA-12 - 10.84%), that situation being caused by the delayed effect of the euro's appreciation which reduced the competitiveness of EA-12 exporters and, respectively, that of the euro zone imports growth above exports growth rate of EA-12, for the second successive year (12.12%). The highest decrease rata of the risk was recorded in 2013 (2.87%) and was influenced by the positively percentage change of the exports of the EU-15 (+1.07%) and by the percentage change of imports EU-15 (-0.26%).

The risk for current account from EU-15 recorded a growth of 1.66% in the 2005-2014 period, compared to the first analysed period and was based, mainly on the increase of the risks from Spain, France, Italy, Greece or Portugal. The balance has been equilibrated by the decrease of financial risks from Germany, Sweden, Austria, Denmark and Netherlands. The position of the member states can be explained, mainly by the evolution of the trade balance and competitiveness of national exporters.

At the EU-15 level, the most performing countries from the perspective of the risk for current account in the first period were Netherlands, France and Belgium, while Greece, Portugal and UK have registered the weakest positions of the indicator. As can be seen in Figure 3, in the second period, the upper trio has changed, following to be composed of Sweden, Germany and Luxembourg, while Spain took the place of UK into the lower trio.

**The risk for external debt service (as a percentage of total exports of goods and services).** Following the examination of the statistical data published by the PRS Group, we found that the minimum point of risk was recorded in 2009 and the maximum in 1995. Given that, the exports decreased in 2009 in EU-15, we can say that debt service coverage ratio fall was caused by the lower debt service, this representing a result of several reductions in interbank interest rate from Economic and Monetary Union (Euribor – 3 months) from 6.82% in 1995 to 1.22% in 2009. The interest rate to which banks borrow in euro on the interbank market recorded this decrease following the reduction of the ECB's monetary policy interest rate and the massive injection of liquidity. The current low risk for external debt service can be explained by low costs with rates, interest rates and commissions (components that can be covered easily from exports revenues), the low level of the risk indicating a high capacity for external debt payment in EU-15.

In 2003, we have witnessed the strongest risk reduction for external debt service in EU-15 (15.95% – caused by the decrease of Euribor – 3 months in 2003 with 2.07 percentage points compared to their level in 2000), while the highest increase was manifested in 2010 (2.96%), the risk analysed recording increases until 2014, this being influenced by the growth of costs with public debt.

In the 2005-2014 period, the risk of external debt manifested in EU-15 decreased by 17.14%, compared to 1995-2004 and was influenced mainly, by the reduction of the interest rate on the interbank market and by liquidity injection.

The first period corresponds to a top 3, composed of Austria, Luxembourg and Ireland, from the point of view of efficient countries in the risk management for external debt service, while the lower ranking includes Greece, Sweden and UK. In the 2005-2014 period, Belgium took the place of Ireland in the upper ranking, while Sweden and UK have improved their position compared to other countries, their former positions in the ranking being taken by Spain and France.

**The risk for exchange rate stability.** In the examined period, the highest level of risk for stability exchange rate took place in 1997, while the minimum level of it was recorded in 2013.

The risk related to exchange rate stability in EU-15 felt a decrease of 6.21% (from 8.944 to 9.501) in the second period, compared to the previous one, this being influenced by the euro adoption in analysed countries, excepting Denmark, UK and Sweden. During the 1995-2004 years, the countries with the most powerful position were UK, Denmark, Germany, Luxembourg and Netherlands. Among the worst performing, we identified countries such as, Sweden, Spain and Italy. The high convergence level of the indicator in EU-15 during 2005-2014 period, excluding the states that have adopted the opt-out clause and Sweden, is explained by the fact that, in this period, the euro was used as a single European currency for the entire analysed period, while this argument only applies for three years from the previous period. At the same time, reducing the risk for exchange rate stability was caused by the application of the EMU principles, oriented to strengthen the euro stability.

**The risk for external debt (as a percentage of GDP).** We identified a maximum point of risk for foreign debt in 2014, while the lowest level of it has been recorded in 1995.

**Figure 3**. *Financial risk components in EU-15 member states*

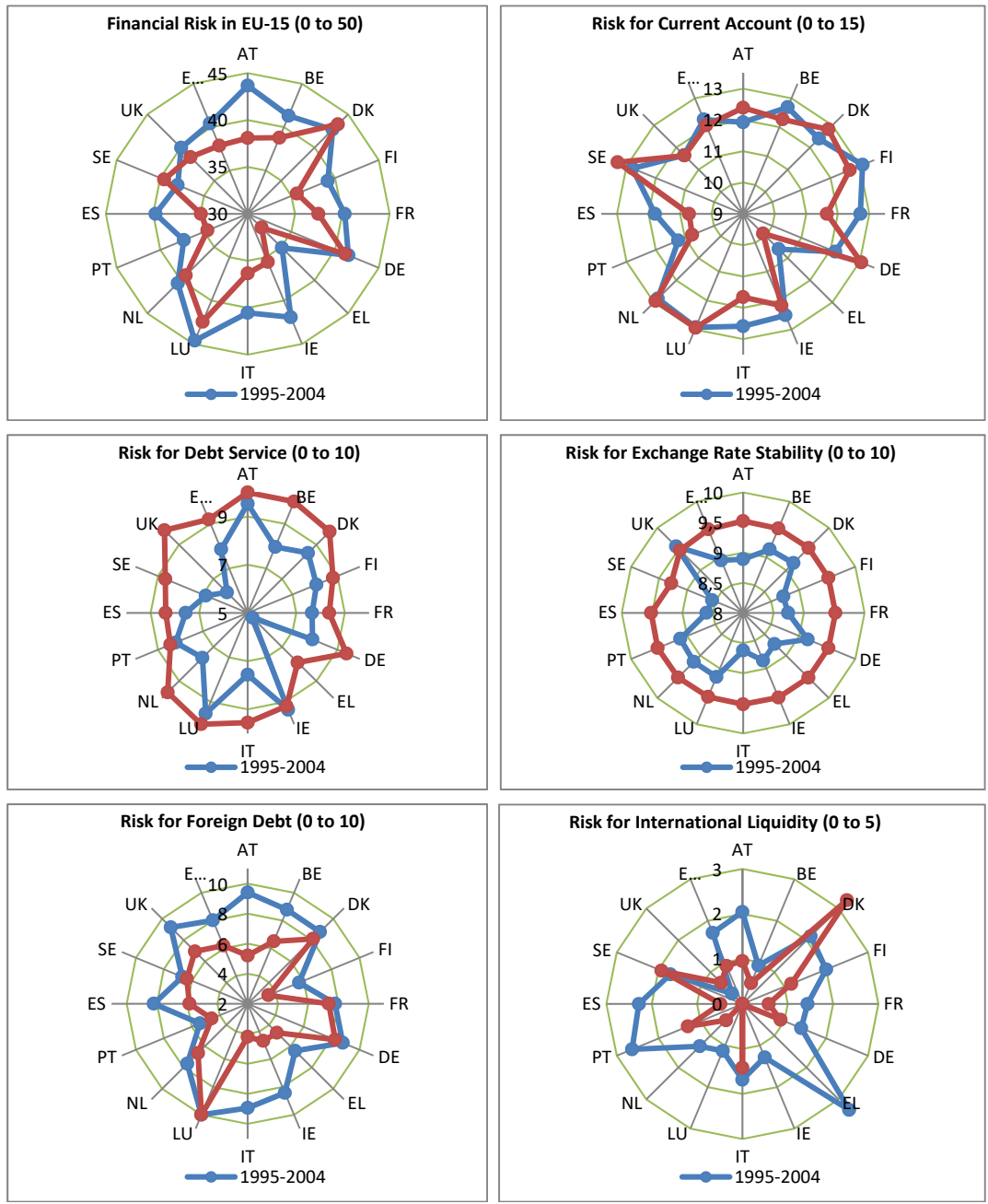

**Source:** Own calculations using PRS Group database.

Thereby, external debt felt an ascending trend and represents one of the biggest problems in the European Union, its unfavorable evolution being mainly caused by the high levels of public debt. This indicator felt the highest growth in 2011 (10.79%), close to the manifestation of the default risk in the euro area (2012), while the risk for external debt has been reduced the most in 2008 (5.33%).

The 2005-2014 period marks an increase of the risk for foreign debt with 22.70% in EU-15, this involution being caused, among other factors, by the high growth of the risk (expressed in deviation points) in Italy (4.74), Austria (4.20) and Ireland (3.75). In the first period, countries such as, Luxembourg, Austria and the UK performed, while the last three positions were held by Greece, Finland and Portugal. In the second period, Germany and Denmark took the place of Austria and UK in the upper ranking and, in terms of the lower ranking, Italy replaced Greece, the increase of the risk for external debt from Italy representing the highest growth from EU-15.

**The risk for international liquidity (expressed as number of months of imports coverage).** The evolution of the risks showed a minimum risk for international liquidity in 1995, covering the necessary imports for a period of about three months, while in 2008 has been recorded the maximum risk level, when official reserves had the capacity to sustain the imports for a period of about 20 days. In this year, the shock of the financial crisis and the inability of banks to grant new loans, in the absence of the recovery of those already made, led to the official reserves utilisation for improving the resilience of the financial system. In 1999, it has occurred the highest percentage increase in risk (34.62%) at the opposite pole being situated the decrease from 2001 (28.54%).

The risk for international liquidity has increased by 46.57% in 2005-2014 compared to 1995-2004, it relying on the increase of the risk (expressed in deviation points) from countries like Greece (3.31), Spain (1.80), respectively Portugal (1.33). Also, there were risk reductions, the most significant being found in Denmark (1.13), UK (0.34) and Sweden (0.21).

At EU-15 level, in the first period, the most performing countries in terms of risk management for international liquidity were UK, Belgium and Luxembourg, the poorly positioned countries being Greece, Portugal and Spain. In the second period, Denmark, Sweden and Italy have entered into the category of countries with low international liquidity risks, while at the opposite pole were countries such as Belgium, Ireland and Luxembourg.

**Financial risk volatility.** In order to capture the volatility of the financial risks in the EU-15, we computed the standard deviation (Figure 4), identifying a level of financial volatility during 1995-2014 of 2.278 deviation points, with a maximum volatility of the risk for external debt for the entire analysed period of 1.114 deviation points, the larger contribution to it being held by the fluctuation of the risk for external debt manifested in 2005-2014 period (0.848 deviation points). Overall, the risk for current account (0.245) and for international liquidity (0.472) recorded the lowest fluctuations, while the risk for external debt service (0.868) and for exchange rate stability (0.681) ensured the balance between the maximum and the minimum deviation from the mean. Thereby, we found that the fluctuation of the risk for external debt had a high impact on the deviation from the mean of the financial risk indicator.

**Model estimation.** In the second phase of the research, we analysed the results of the estimated model, starting from the investigation of the data series characteristics.

Following the examination of the stationarity (Table 1), we found that all used tests confirm this hypothesis on the following variables: economic growth and net taxes on products, respectively economic growth, budget balance, real long-term interest rate, inflation rate, financial risk and risk for economic growth, the last ones being lagged by 1 year.

The only variable on which half of the tests argued the presence of a unit root and half confirmed its stationarity is the variable related to import (expressed as a percentage of GDP). In this regard, we processed correlogram to verify the autocorrelation function and we found the presence of the trend and constant, because for 12 lags considered, there is a trend of decline in function. Taking into account this feature, we confirmed the stationarity of the variable because, in that case, all tests confirmed the alternative hypothesis. Thereby, it can be stated that, the variables included in the model have been fluctuating around average, which confirms the previously assumed method (least squares).

**Figure 4**. *Financial risk volatility in EU-15 (Standard deviation)*

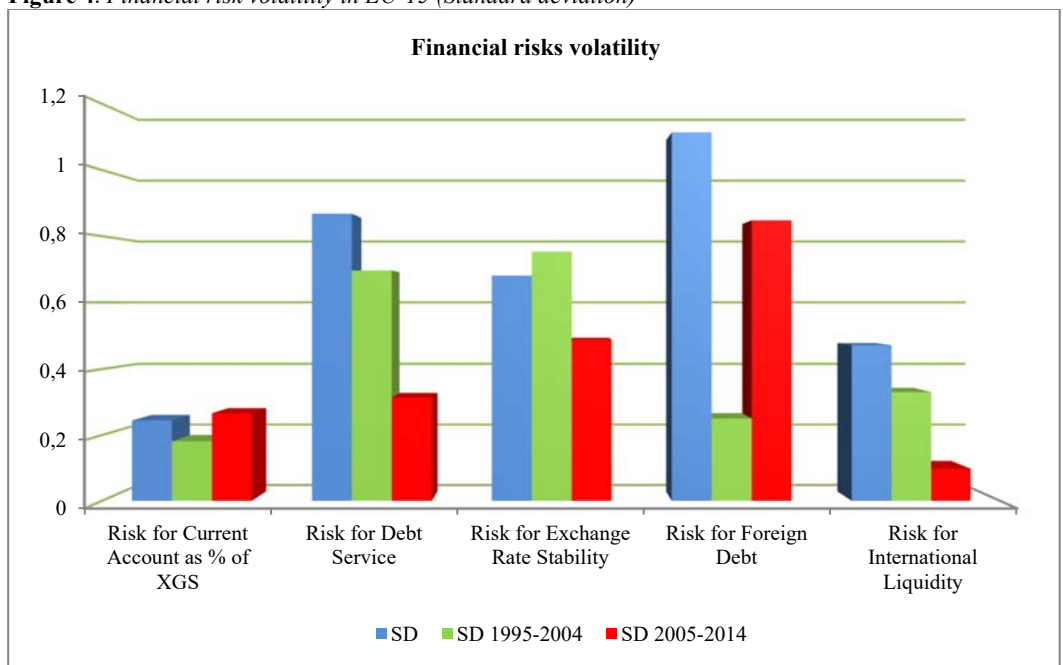

**Source:** Own calculations using PRS Group database.

**Table 1.** *Stationarity of the variables at first order of integration - I(0)*

| Variable | Levin, Lin and Chu | | | Breitung | Im, Pesaran and Shin | | ADF-Fisher | | | PP-Fisher | | |
|---|---|---|---|---|---|---|---|---|---|---|---|---|
| | * | ** | *** | * | * | ** | * | ** | *** | * | ** | *** |
| y | 0.00 | 0.00 | 0.00 | 0.00 | 0.00 | 0.00 | 0.00 | 0.00 | 0.00 | 0.00 | 0.00 | 0.00 |
| y(-1) | 0.00 | 0.00 | 0.00 | 0.00 | 0.00 | 0.00 | 0.00 | 0.00 | 0.00 | 0.00 | 0.00 | 0.00 |
| imp | 0.00 | 0.01 | **1.00** | 0.00 | 0.00 | **0.68** | 0.00 | **0.61** | **1.00** | 0.01 | **0.80** | **1.00** |
| npt | 0.02 | 0.00 | **0.83** | 0.00 | 0.00 | 0.00 | 0.00 | 0.00 | **1.00** | 0.03 | 0.00 | **1.00** |
| budget(-1) | 0.00 | 0.00 | 0.00 | 0.01 | 0.00 | 0.00 | 0.00 | 0.00 | 0.00 | 0.01 | 0.00 | 0.00 |
| intrate(-1) | 0.00 | 0.00 | 0.00 | **0.39** | 0.00 | **0.07** | 0.00 | 0.03 | 0.00 | 0.00 | **0.06** | 0.00 |
| inflation(-1) | 0.00 | 0.00 | 0.00 | 0.00 | 0.00 | 0.00 | 0.00 | 0.00 | 0.00 | 0.00 | 0.00 | 0.00 |
| finrisk(-1) | 0.00 | 0.00 | 0.00 | **0.07** | 0.00 | 0.00 | 0.00 | 0.00 | 0.01 | **0.73** | **0.06** | 0.00 |
| yrisk(-1) | 0.00 | 0.00 | 0.02 | **0.28** | 0.00 | **0.04** | 0.00 | 0.03 | **0.81** | 0.00 | 0.02 | **0.94** |

*trend and constant; ** constant; *** absence of trend and constant.
**Source:** Own calculations using Eviews 9.0.

Prior to estimate the model, we had to test its compatibility with random effects or fixed effects method by performing the Hausman test (Correlated Random Effects) and the Redundant Fixed Effects – Likelihood Ratio test. Given that the probability associated with the test Hausman (0%) is less than 5%, we rejected the null hypothesis which assumes that the model is compatible with the random effects method and accepted the alternative one, that indicates the use of the fixed effects method, as being more appropriate. For the avoidance of doubt on the correct selection of the estimation method, we ran the Redundant Fixed Effects – Likelihood Ratio test, its null hypothesis consisting in the rejection of the compatibility with the fixed effects method. Given that the probability of the test (0%) is less than 5%, we rejected the null hypothesis and accepted the alternative one, confirming that the model is compatible with the fixed effects method.

Further, we estimated the regression using the fixed effects method and analysed the obtained results (Annex 2). The coefficient of determination (R-squared) shows that the used exogenous variables explain 90.15% of the endogenous variable fluctuation, which indicates that we used a correct selection of the explanatory factors. The probabilities associated with the explanatory variables prove that, for each estimator, the probability of the coefficients to be insignificant is less than 1%. Regarding the coefficients examination, we analysed the estimated impacts respecting the „ceteris paribus" condition.

The results of the estimated regression prove the negative impact of financial risks on economic growth from the next year, an increase (decrease) by 1 deviation point of the indicator, which implies a decrease (increase) of financial risks – considering the inverse scaling of the indicator –, leading to an modification of GDP growth rate of +0.192 (-0.192) percentage points. The negative impact of the increase of the financial risks on economic growth in the EU-15 was caused by the manifestation of the risks of its components, as well as by the new feelings of fear and panic, recorded by the economic agents as a result of financial disturbances experienced during the crisis. For a better understanding of the impact of these 5 types of financial risks on economic growth, we explained their effects in Figure 5.

**Figure 5.** *Financial risks effects on economic growth*

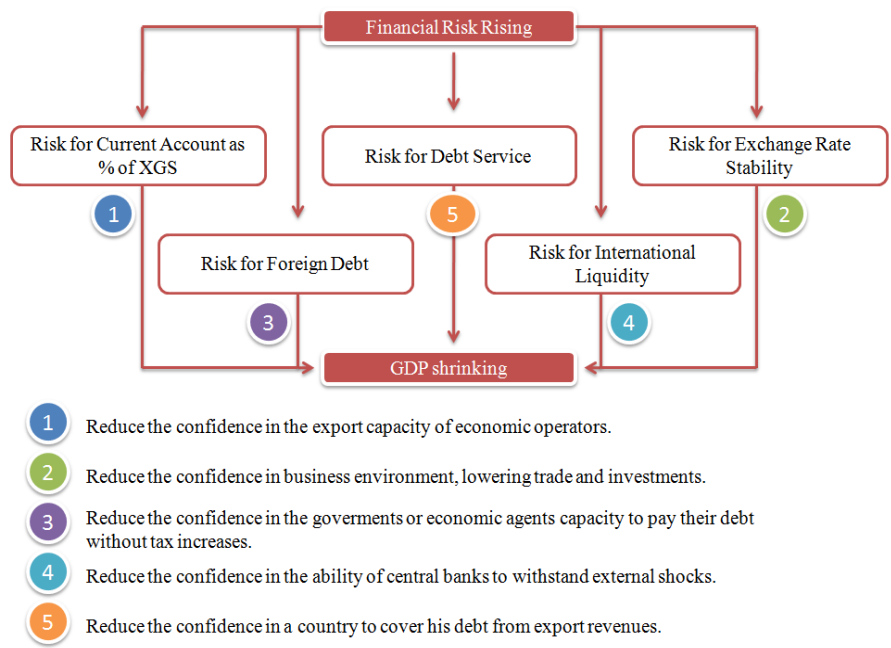

**Source:** Own calculations using Microsoft Office PowerPoint 2007.

According to the results attached in the Annex 2, an increase of the economic growth rate in the previous year by one percentage point lead to a change in current economic growth with 0.536 percentage point, that being caused by the manifestation of the optimism and by the presence of an great expectations behaviour from the economic operators. The model estimated a negative impact of 0.027 percentage points of an increase in imports by 1 percentage point (expressed as a percentage of GDP), on the economic growth and a positive effect on the percentage change in real GDP of 0.166 percentage points, due to the increase of the net taxes on products by one percentage point of GDP. These two reactions has derived from the formulas related to GDP calculation methods (expenditures approach and production approach).

The increase (decrease) of budget balance (expressed as a percentage of GDP) by 1 percentage point from the previous year has a positive (negative) impact of 0.095 percentage points on the economic growth rate in EU-15, that being influenced mainly by the orientation of the Member States to counter-cyclical economic policies adoption and by the impact of rising budget deficit on the public debt growth, also affecting interest rate and investment. However, 1 percentage point change of the variable related to real long-term interest rate lagged by one year, resulted in a rate of economic growth, lower by 0.207 percentage points. On the other hand, the reaction of economic growth as a result of a 1 percentage point change in consumer price index from the previous year, was negative (-0.896 percentage points), as a consequence of its impact on reducing consumption. A last control variable analysed consists in the risk for economic growth from the previous year. We found that an increase (decrease) in its evolution by 1 deviation points, which implies a decrease (increase) of the risk, lead to a reduction (increase) of economic growth in the current year with 0.337 percentage points. This reaction is influenced by the tendency of governments to accumulate buffers in times with low risks and to release them in periods where there is manifested an intensification of them. If all variables remain constant, economic growth changes by 2,255 percentage points, indicating that there are still other variables not included in the model, which may impact the level of GDP.

Regarding the confirmation of the model validity, we used F-stat test, whose value confirms this hypothesis (109.094) since it is greater than F-critical (1.661) and its likelihood (0%) is lower than the significance threshold of 5%.

Next, we checked the assumptions required for the model validation, test results used for this purpose, being attached in Table 2. Jarque-Berra test confirmed the absence of arguments to reject the assumption that the residuals are normally distributed, taking into consideration the fact that the associated probability (7.8%) is greater than 5%, while the Durbin-Watson test (2.094) is not conclusive to confirm or to reject the autocorrelation of the residuals, since DW value is greater than 4-DU and less than 4-DL. Despite the fact that, the Durbin-Watson test recommended accepting the hypothesis that autocorrelation is negative, in order to provide an exhaustive analysis of the autocorrelation phenomenon, we further analysed the corresponding probabilities of Breusch-Pagan LM (100%) and Pesaran CD (83.6%) tests, which confirmed the absence of cross-section dependence, given that their likelihood is higher than the threshold of 5%. Regarding homoscedasticity, we attached the explanations box from Table 2.

An other hypothesis to be confirmed for validating the results of the model consists in the absence of the multicollinearity. In order to test the presence of multicollinearity, we used the Klein's criterion (1962), a method which supports the presence of the multicollinearity hypothesis only if the correlations between the regressors are greater than the coefficient of determination. Table 3 proves the absence of the multicollinearity, since no correlation between regressors does not exceed 90%.

Taking into consideration the hypotheses tested, we can assume that our analysis is valid and meets all the required criteria for a correct specified econometric model.

**Table 2.** *Checking the hypotheses for model validation*

| Hypothesis tested | Test | Probability / Result | Hypothesis accepted |
|---|---|---|---|
| Compatibility with random effect model | Correlated Random Effects - Hausman Test | 0.000 | NO |
| Compatibility with fixed effect model | Redundant Fixed Effects - Likelihood Ratio | 0.000 | YES |
| Normal distribution of the residuals | Jarque-Berra | 0.078 | YES |
| Absence of residuals dependence** | Breusch-Pagan LM | 1.000 | YES |
| Absence of residuals dependence** | Pesaran CD | 0.836 | YES |
| Homoscedasticity*** | White | 256.950** | YES |

*We attached the histogram in Figure 6; ** at cross-sections level; *** White test result is less than Chi-squared statistic for 266 degrees of freedom (305.041) -> n-k-1 (285 observations, 9 regressors and 9 dummy variables, included in fixed effect model), which led to the acceptance of homoscedasticity.
**Source:** Own calculations using Eviews 9.0.

**Figure 6.** *Residuals histogram*

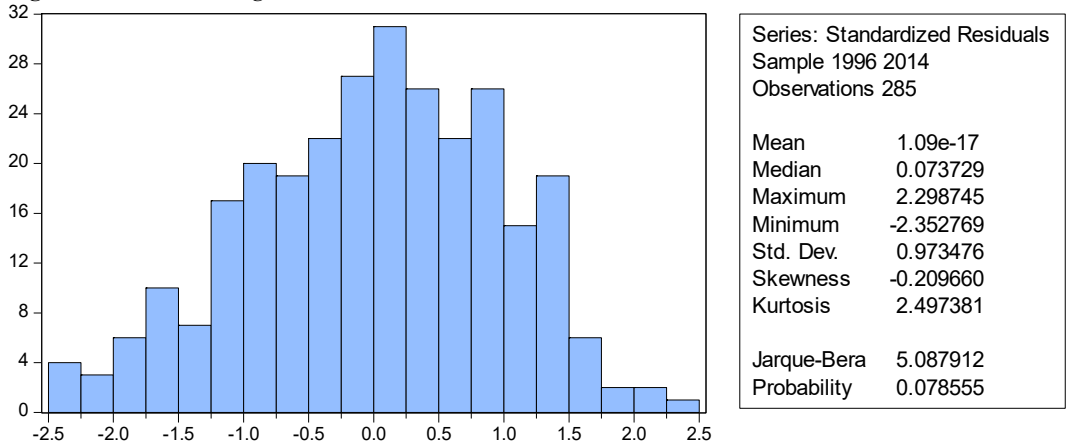

Source: Own calculations using Eviews 9.0.

**Table 3.** *Checking for multicollinearity-Klein's criterion*

| Correlation matrix | Y* | IMP | NPT | Budget* | Intrate* | Inflation* | Finrisk* | Yrisk* |
|---|---|---|---|---|---|---|---|---|
| Y* | **1.00** | 0.17 | -0.01 | 0.44 | -0.25 | 0.20 | 0.31 | 0.68 |
| IMP | 0.17 | **1.00** | -0.09 | 0.26 | -0.19 | -0.04 | 0.26 | 0.13 |
| NPT | -0.01 | -0.09 | **1.00** | 0.27 | -0.09 | -0.09 | 0.02 | 0.01 |
| Budget* | 0.44 | 0.26 | 0.27 | **1.00** | -0.38 | -0.03 | 0.32 | 0.63 |
| Intrate* | -0.25 | -0.19 | -0.09 | -0.38 | **1.00** | -0.14 | -0.06 | -0.44 |
| Inflation* | 0.20 | -0.04 | -0.09 | 0.03 | -0.15 | **1.00** | -0.09 | 0.23 |
| Finrisk* | 0.31 | 0.26 | 0.02 | 0.32 | -0.06 | -0.09 | **1.00** | 0.17 |
| Yrisk* | 0.68 | 0.13 | 0.01 | 0.63 | -0.44 | 0.23 | 0.17 | **1.00** |

R-squared = 0.90; *Variables lagged by 1-year.
**Source:** Own calculations using Eviews 9.0.

## 5. Conclusions

Our analysis shows that in the EU-15 were accumulated excessive financial risks in the 2005-2014 period, relying mainly on the increase of the risks for external debt, current account or international liquidity. The financial risk has reached the highest level in countries like Greece, Portugal and Spain and had severe adverse consequences on economic growth in those states. Surprisingly, the risk for external debt service has experienced a downward trend, which was caused by the lowering of the interbank interest rates.

The data source used reinforces the EU-15 dichotomy in winners and losers of the European project and demonstrates the existence of financial imbalances at European Union level, one of these consisting in the effect of increasing the export share of Germany in Portugal and Greece on the growth of financial risks recorded as well as on economic growth.

The estimated model demonstrated a negative influence of an increase in EU-15 financial risks on economic growth, impact that can be transmitted through the effect of declining the confidence in the country to face financial challenges and through the intrinsic effect. The coefficient is significant and the tests performed confirmed the validity of the model, which gives certainty to our model.

It is clear that the EU-15 is divergent, even in terms of accumulation of financial risks, the risk reduction in a country, offering the premises for an increase of risks in other countries, through commercial, financial and political channel.

**Acknowledgements**

Through this research, we want to express our gratitude to Professor Dorel Ailenei, former Dean of the University of Economic Studies, Faculty of Theoretical and Applied Economics, for the moments when he supported and encouraged us in our first steps made in the economic research area.

His extensive background in economic policy, regional economy and econometrics have helped us to strengthen our economic knowledge and to apply them in the conducted analyses. His enthusiasm and economic research oriented spirit filled the souls of thousands of students with the desire to perform in economic research and will be reflected forever in our work, regardless of the time and place where we find ourselves.

**Annex 1.** Structure of the model

| Variable | Unit | Source |
| --- | --- | --- |
| **Endogenous** | | |
| Real Gross Domestic Product | Percentage change | Eurostat |
| **Exogenous** | | |
| Imports | Percentage of GDP | Eurostat |
| Budget balance | Percentage of GDP | AMECO |
| Real long-term interest rate | Rate | AMECO |
| Inflation rate | Percentage change of consumer prices | World Bank |
| Net taxes on products | Percentage of GDP | Eurostat |
| Financial risk | Index (risk rating) | PRS Group |
| Risk for economic growth | Index (risk rating) | PRS Group |

**Source:** Own calculations.

**Annex 2.** Model estimation

Dependent Variable: Y
Method: Panel EGLS (Cross-section SUR)
Sample (adjusted): 1996 2014
Periods included: 19
Cross-sections included: 15
Total panel (balanced) observations: 285
Linear estimation after one-step weighting matrix

| Variable | Coefficient | Std. Error | t-Statistic | Prob. |
| --- | --- | --- | --- | --- |
| Y(-1) | 0.536103 | 0.024134 | 22.21362 | 0.0000 |
| IMP | -0.026784 | 0.006572 | -4.075655 | 0.0001 |
| NPT | 0.166397 | 0.053833 | 3.091008 | 0.0022 |
| BUDGET(-1) | 0.094992 | 0.011775 | 8.067331 | 0.0000 |
| INTRATE(-1) | -0.207414 | 0.019183 | -10.81225 | 0.0000 |
| INFLATION(-1) | -0.895764 | 0.031996 | -27.99620 | 0.0000 |
| FINRISK(-1) | 0.192455 | 0.011514 | 16.71498 | 0.0000 |
| YRISK(-1) | -0.336776 | 0.026335 | -12.78798 | 0.0000 |
| C | -2.255361 | 0.790613 | -2.852672 | 0.0047 |

Effects Specification

Cross-section fixed (dummy variables)

Weighted Statistics

| | | | |
| --- | --- | --- | --- |
| R-squared | 0.901580 | Mean dependent var | 0.428365 |
| Adjusted R-squared | 0.893316 | S.D. dependent var | 3.146942 |
| S.E. of regression | 1.013523 | Sum squared resid | 269.1342 |
| F-statistic | 109.0942 | Durbin-Watson stat | 2.094182 |
| Prob(F-statistic) | 0.000000 | | |

**Source:** Own calculations using Eviews 9.0.